\documentclass{WileyMSPtemplate}
\pdfoutput=1
\usepackage[labelfont=bf]{caption}
\usepackage{setspace}
\usepackage{graphicx}
\graphicspath{ {./images/} }
\usepackage[numbers]{natbib}
\usepackage[superscript,biblabel]{cite}
\usepackage{hyperref}

\begin{document}
\begin{sloppypar}

\pagestyle{fancy}

\title  {Kinetic Insights into Bridge Cleavage Pathways in Periodic Mesoporous Organosilicas}

\maketitle

\author{Zeming Sun}
\author{Aine Connolly}
\author{Michael O. Thompson*}

\begin{affiliations}
Dr. Z. Sun, A. Connolly, Prof. M. O. Thompson\\
Materials Science and Engineering, Cornell University, Ithaca, New York 14853, USA\\
Email Address: mot1@cornell.edu (M.O.T.)
\end{affiliations}


\begin{abstract}
Bridging functionalities in periodic mesoporous organosilicas (PMOs) enable new functionalities for a wide range of applications. Bridge cleavage is frequently observed during anneals required to form porous structures, yet the mechanism of these bridge cleavages has not been completely resolved. Here, we reveal these chemical transformations and their kinetic pathways on sub-millisecond timescales induced by laser heating. By varying anneal times and temperatures, the transformation dynamics of bridge cleavage and structural transformations, and their activation energies, are determined. The structural relaxation time for individual reactions and their effective local heating time are determined and compared, and results directly demonstrate the manipulation of different molecules through kinetic control of the sequence of reactions. By isolating and understanding the earliest stage of structural transformations, this study identifies the kinetic principles for new synthesis and post-processing routes to control individual molecules and reactions in PMOs and other material systems with multi-functionalities.
\end{abstract}

\section{Introduction}

\begin{figure}[tb!]
\centering
\includegraphics[width=\linewidth]{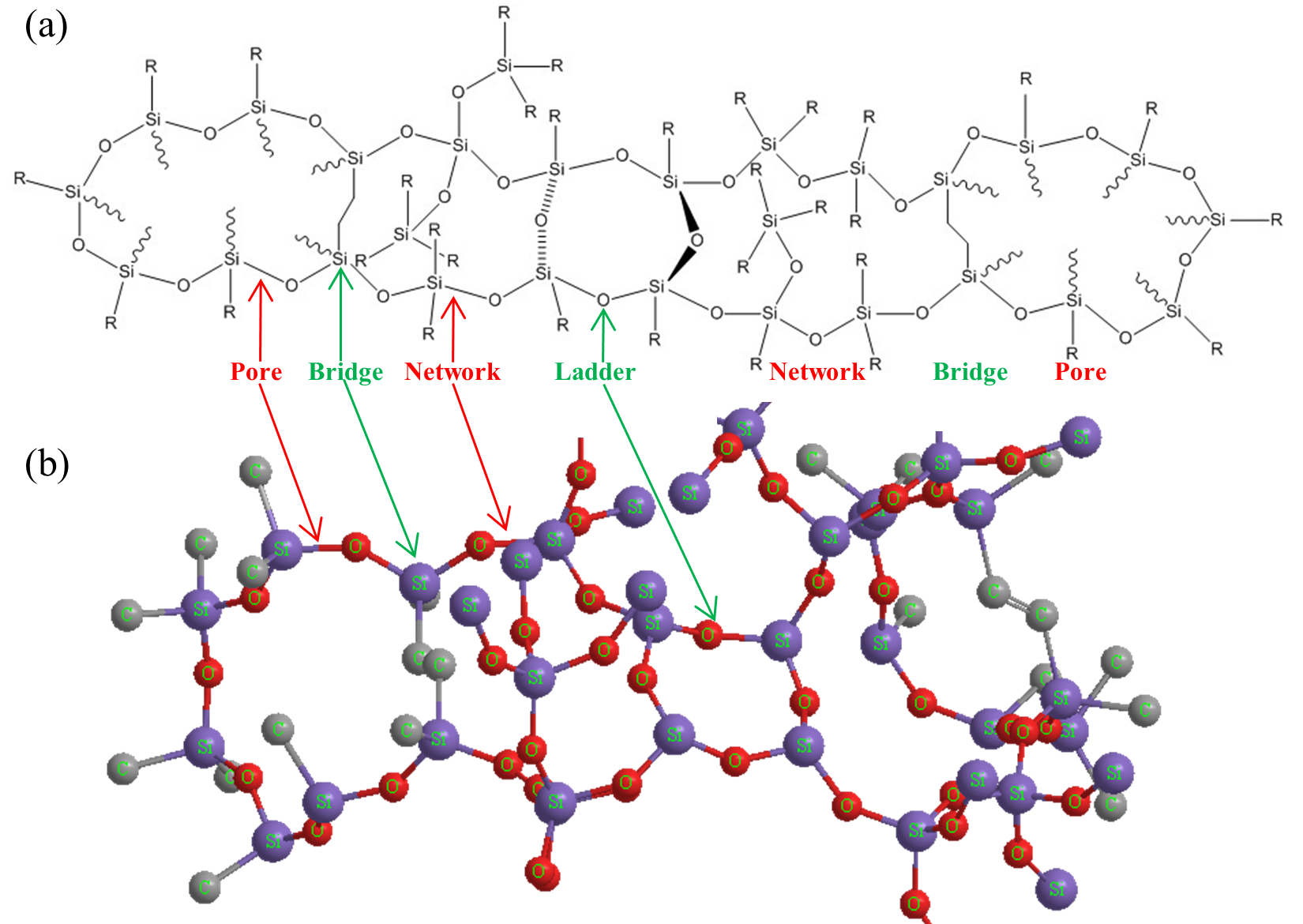}
\caption{(a) 2D chemical structure and (b) 3D projection of ethyl-bridged organosilicas, showing the intrinsic carbon-bridges in addition to pore, network, and ladder motifs.}
\label{SunFig1}
\end{figure}

Periodic mesoporous organosilicas (PMOs), porous silica networks containing organic entities, have attracted tremendous attention both for their well-ordered porous structures with controlled sizes, and for the interactions with molecules, atoms, and ions enabled by tailored bridge structures \cite{SunRef1}. Since the discovery of ordered materials \cite{SunRef2} and periodic mesoporous organosilicas \cite{SunRef3,SunRef4,SunRef5}, their distinctive features have found use in a wide variety of applications including catalysis \cite{SunRef6,SunRef7}, bioengineering and drug delivery \cite{SunRef8}, low-$\textit{k}$ dielectrics \cite{SunRef9}, adsorption/separation of metals and gases \cite{SunRef6,SunRef8}, ion exchanges \cite{SunRef10}, and optics \cite{SunRef8}. 

\bigbreak

Sol-gel synthesis \cite{SunRef11,SunRef12} of organic-inorganic PMOs have been widely investigated to develop a range of functionalities, synthesis routes, and properties \cite{SunRef6,SunRef8,SunRef10,SunRef13}. Advances include introducing specific functional bridge components, from simple chain bridges (methyl \cite{SunRef14}, ethyl \cite{SunRef4,SunRef5,SunRef15,SunRef16,SunRef17,SunRef18,SunRef19,SunRef20,SunRef21,SunRef22}, and ethene \cite{SunRef3,SunRef4,SunRef23,SunRef24,SunRef25}) to more complicated bridges such as benzene \cite{SunRef23,SunRef26,SunRef27,SunRef28}, thiophene \cite{SunRef29,SunRef30}, and ring structures \cite{SunRef31}, as summarized in \textbf{Table~\ref{tbl:1}}. The bridge scheme has been further extended to incorporate non-carbon atoms (\textit{e.g.}, S, Al, N) into bridges using either co-condensation \cite{SunRef32,SunRef33,SunRef34,SunRef35} or post-grafting \cite{SunRef36} methods, and even expanded to non-silica-based mesoporous materials including metal-oxide-based \cite{SunRef37} and sulfide-based \cite{SunRef38} materials.

\bigbreak

The functional bridges are of critical interest in PMOs due to the varied functionalities establishing wide-ranging chemical properties \cite{SunRef6,SunRef8,SunRef10,SunRef39}. Moreover, additional bridge connections within the framework provide opportunities for tuning mechanical, dielectric, and optical characteristics \cite{SunRef8,SunRef40}. Additionally, the integration of these organic bridge motifs can tailor pore structures for surface properties at the nano-scale \cite{SunRef21}.

\bigbreak

However, bridges in PMOs are thermally unstable and may undergo substantial undesirable cleavage during either synthesis or post heat treatment \cite{SunRef6,SunRef41}. Several representative bridge-cleavage cases using varied bridge structures, templates, synthesis conditions, and post heat treatments are summarized in Table~\ref{tbl:1}. Factors inducing bridge cleavages include annealing temperatures, acid or base pH levels, extraction steps, processing environment, functionality sizes, and pore sizes.

\begin{table} [b!]
  \caption{Carbon-bridges cleavage conditions for periodic mesoporous organosilicas.}
  \label{tbl:1}
  \scalebox{0.86}{
  \begin{tabular}{p{2.5cm} p{1.5cm} p{6cm} p{2.5cm} p{3cm} p{0.5cm}}
    \hline
    Bridges (Precursors) & Templates & Sol-gel synthesis & Post heat treatment & Bridge cleavage & Ref. \\
    \hline
    Si-CH$_2$-Si (BTEM\textsuperscript{\emph{a}}) & CTAB, CTAC & Aging (25\,\textdegree C, 96\,h) & 400\,--\,700\,\textdegree C, 4\,h, in air & Onset at 400\,\textdegree C; completion at 600\,\textdegree C in air & \cite{SunRef14} \\
   Si-CH$_2$CH$_2$-Si (BTME\textsuperscript{\emph{b}} or BTEE\textsuperscript{\emph{c}}) & CTAB, CTAC, OTAC & Multiple stirring (25\,\textdegree C, $>$\,24\,h), aging (95\,\textdegree C, 21\,h)\,$+$\,extraction (50\,\textdegree C, 6\,h) & TGA & 400\,\textdegree C in nitrogen; 300\,\textdegree C in air & \cite{SunRef5} \\
    & C$_n$TMACl & Multiple stirring (25\,\textdegree C, $>$\,24\,h), aging (95\,\textdegree C, 21\,h)\,$+$\,extraction (50\,\textdegree C, 6\,h) & TGA & 540\,\textdegree C in nitrogen; 280\,\textdegree C in air & \cite{SunRef16} \\
   & Brij 76, Brij 30 & Stirring (50\,\textdegree C and 25\,\textdegree C, $>$\,18\,h), aging (95\,\textdegree C, 24\,h)\,$+$\,extraction (50\,\textdegree C, 8\,h) & TGA & 400\,\textdegree C in nitrogen; 338\,\textdegree C in air & \cite{SunRef17} \\
   & P123 & Stirring (35\,\textdegree C, 24\,h), aging (85\,\textdegree C, 24\,h)\,$+$\,extraction (25\,\textdegree C, 10\,h) & 50\,--\,800\,\textdegree C, 17\,h, in air & 175\,\textdegree C in air &  \cite{SunRef19} \\
   &&$-$& 35\,--\,300\,\textdegree C, in air & 200\,--\,300 \textdegree C in air & \cite{SunRef20} \\
   & PEO-PLGA-PEO & Stirring (25\,\textdegree C, 0.5\,h), aging (100\,\textdegree C, 24\,h)\,$+$\,extraction (25\,\textdegree C) & 500\,\textdegree C, 10\,h, in nitrogen & 500\,\textdegree C in nitrogen & \cite{SunRef22} \\
   Si-CH$=$CH-Si (BTEENE\textsuperscript{\emph{d}}) & CTAB & Aging (25\,\textdegree C, 96\,h) & 40\,--\,400\,\textdegree C, 6\,h, in air & Onset at 350\,\textdegree C; completion at 400\,\textdegree C in air & \cite{SunRef3} \\
   & Brij 76 & Stirring (50\,\textdegree C, 12\,h), aging (90\,\textdegree C, 24\,h)\,$+$\,extraction (25\,\textdegree C) & 200\,--\,1000\,\textdegree C, 4\,h, in air & 200\,--\,350\,\textdegree C in air & \cite{SunRef23} \\
   & P123 & Stirring (40\,\textdegree C, 24\,h), aging (100\,\textdegree C, 24\,h)\,$+$\,extraction & 40\,--\,400\,\textdegree C, in air & 320\,\textdegree C in air & \cite{SunRef25} \\ 
  Si-C$_6$H$_4$-Si (BTEB\textsuperscript{\emph{e}}) & OTAC & Aging (95\,\textdegree C, 20\,h) & TGA & 500\,\textdegree C in air or nitrogen & \cite{SunRef26} \\ 
  & C$_n$TMACl & Stirring (25\,\textdegree C, 24\,h), aging (92\,\textdegree C, 24\,h)\,$+$\,extraction (60\,\textdegree C, 6\,h) & TGA & 500\,\textdegree C in air or nitrogen & \cite{SunRef27} \\
  & Brij 76 & Stirring (50\,\textdegree C, 12\,h), aging (90\,\textdegree C, 24\,h)\,$+$\,extraction (25\,\textdegree C) & 200\,--\,1000\,\textdegree C, 4\,h, in air & 600\,\textdegree C in air & \cite{SunRef23} \\
   & P123 & Stirring (39\,\textdegree C, 20\,h), aging (100\,\textdegree C, 24\,h)\,$+$\,extraction (100\,\textdegree C, 72\,h) & 250\,\textdegree C, in air; TGA & 550\,\textdegree C in air & \cite{SunRef28} \\
  Si-C$_4$H$_4$S-Si (BTET\textsuperscript{\emph{f}}) & P123 & Stirring (40\,\textdegree C, 20\,h, aging (100\,\textdegree C, 24\,h)\,$+$\,extraction (25\,\textdegree C, 6\,h) & TGA & 400\,\textdegree C in air & \cite{SunRef29} \\
  & PEO-PLGE-PEO & Stirring (40\,\textdegree C, 24\,h), different acidity, aging (100\,\textdegree C, 24\,h)\,$+$\,extraction (56\,\textdegree C, 24\,h) & TGA & 300\,--\,450\,\textdegree C in mixed air and nitrogen & \cite{SunRef30} \\
  Carbon rings ([(EtO)$_2$-SiCH$_2$]$_3$) & CTAB & Stirring (20\,\textdegree C, 24\,h, aging (80\,\textdegree C, 24\,h\,$+$\,extraction with stirring (25\,\textdegree C, 48\,h) & 300\,--\,700\,\textdegree C, in nitrogen & 400\,\textdegree C in nitrogen & \cite{SunRef31} \\
  Block copolymers & PEO-PLGA-PEO & Stirring (25\,\textdegree C, 3\,h and 40\,\textdegree C, 1\,h), aging (95\,\textdegree C, 24\,h)\,$+$\,extraction (25\,\textdegree C, 2\,h), stirring (95\,\textdegree C, 5\,h), drying (100\,\textdegree C, 24\,h) & TGA & 400\,\textdegree C in air & \cite{SunRef33} \\
  Heteroatom cocondensation $[\mathrm{N-, S-}]$ & CTAC & Stirring (25\,\textdegree C, $>$\,12\,h), drying (60\,\textdegree C, 12\,h)\,$+$\,extraction (25\,\textdegree C, 6\,h), drying (60\,\textdegree C, 10\,h under vacuum) & TGA & 300\,\textdegree C (no extraction); 200\,\textdegree C (after extraction) & \cite{SunRef34} \\
  $[\mathrm{Al-}]$ isopropoxide & CTAB & Stirring (25\,\textdegree C, 20\,h), aging (100\,\textdegree C, 24\,h)\,$+$\,extraction (25\,\textdegree C, 6\,h) & TGA & 300\,\textdegree C in air & \cite{SunRef35} \\
   \hline
\end{tabular}
}
\textsuperscript{\emph{a}}Bis(triethoxysilyl)methane;
  \textsuperscript{\emph{b}}Bis(trimethoxysilyl)ethane;
  \textsuperscript{\emph{c}}Bis(triethoxysilyl)ethane;
  \textsuperscript{\emph{d}}Bis(triethoxysilyl)ethylene;
  \textsuperscript{\emph{e}}Bis(triethoxysilyl)benzene;
  \textsuperscript{\emph{f}}Bis-(triethoxysilyl)thiophene.
\end{table}

\bigbreak

During synthesis, control of pH is critical to minimize bridge cleavage. The multiple-hour stirring and aging under controlled pH conditions allow hydrolysis and condensation reactions to occur in a transient state without affecting bridge components, resulting in minimal bridge cleavage for synthesis temperatures below 100\,\textdegree C \cite{SunRef18,SunRef24,SunRef32}.  Complicated and large bridges, however, are challenging to stabilize in the typically used strong acid or base conditions \cite{SunRef42}. 

\bigbreak

Once the framework is established during synthesis, two approaches are commonly exploited to remove templates (\textit{i.e.}, porogens) and create porous structures. Low-temperature chemical extraction is able to avoid loss of bridge moieties inside the framework \cite{SunRef16,SunRef17,SunRef18,SunRef19,SunRef22,SunRef24,SunRef26,SunRef27,SunRef28}, while the calcination approach, a post heat treatment for mesoporous silica class materials \cite{SunRef22}, is challenging \cite{SunRef16,SunRef19,SunRef23,SunRef28}. Thermal gravimetric analyses (TGA) show that it is difficult to protect organic bridges even at moderate calcination temperatures (200\,--\,350\,\textdegree C) for minute time frames. At higher temperatures ($>$\,400\,\textdegree C), the thermal stability of these materials is a significant issue.

\bigbreak

After synthesis, PMOs must also tolerate processing conditions required for many electronic devices and other applications, $\textit{i.e.}$, heating to temperatures above 400\,\textdegree C \cite{SunRef6,SunRef43}. Consequently, although extraction approaches may initially remove templates without compromising bridges, these bridges may not survive subsequent processing \cite{SunRef4,SunRef34}. In most cases, the bridges must be preserved for their functionalities, while some applications may benefit from the generation of terminal dangling bonds of broken bridges \cite{SunRef9,SunRef13,SunRef14,SunRef31}. 

\bigbreak

To pursue either of the goals, understanding the underlying kinetics of bridge cleavages is critical and the focus of this work. By determining activation energies and transformation kinetics of each inorganic/organic motif, this work provides guidance to develop heat treatments for PMOs which would facilitate their transition to widespread applications. This work also potentially provides alternative thermal routes to replace the current pH-dependent synthesis protocols that are time-consuming, relatively unstable, and difficult for large bridge structures. 

\bigbreak

In literature (Table~\ref{tbl:1}), the onset of bridge cleavage is reported to start between 200\,\textdegree C and 600\,\textdegree C in the air; the variations may be related to variation in processing atmosphere, bridge type, and template type \cite{SunRef6}. Several researchers report thermal stability increases by 60\,--\,260\,\textdegree C heating in nitrogen rather than air \cite{SunRef5,SunRef16,SunRef17}, while other researchers report no significant changes with atmosphere \cite{SunRef26,SunRef27}. Reduced temperature stability in the air may be a result of reactions between oxygen and carbon bridges \cite{SunRef6,SunRef16,SunRef19,SunRef29,SunRef34}. However, water generated by ongoing condensation during post heat treatment \cite{SunRef44} may be another source for bridge attacks together with possible nitrogen reactions \cite{SunRef16,SunRef17,SunRef34}; this behavior could explain the lack of significant ambient effects. Some research even suggests the onset temperature for bridge cleavages in nitrogen is lower than in air \cite{SunRef23,SunRef28}. Although the influence of the ambient gas is not fully clear, we conclude, with caution, that heating ambients are a secondary effect influencing bridge cleavage, with temperature dynamics, as investigated in this work being the primary influence.  

\bigbreak

The specific chemical structure of the bridge and template components may affect the overall behavior. For example, benzene bridges show slightly higher bridge-cleavage temperatures than other bridges (Table~\ref{tbl:1}). Templates primarily impact the pore size rather than bridge stability; in ethyl-bridged materials, CTAB ($\sim$\,2\,nm pore), Brij-76 ($\sim$\,5\,nm pore), P123 ($\sim$\,15\,nm pore), and PEO-PLGA-PEO (thicker pore wall) templates show similar cleavage temperatures at $\sim$\,300\,\textdegree C in air. In this work, we focus on the Brij-76 template which yields medium-size pores, and ethyl-bridged organosilicas (\textbf{Figure~\ref{SunFig1}}). This combination has been widely used to make PMO products (Table~\ref{tbl:1}), and the general behavioral characteristics can be extended for use in a wide variety of block \cite{SunRef32,SunRef33,SunRef34,SunRef35} and graft \cite{SunRef36} copolymers.

\bigbreak

Studies of thermal decomposition are generally based on either TGA (weight loss as a function of temperature) \cite{SunRef3,SunRef5,SunRef14,SunRef15,SunRef16,SunRef17,SunRef20,SunRef23,SunRef25,SunRef26,SunRef27,SunRef28,SunRef29,SunRef30,SunRef31,SunRef33,SunRef34,SunRef35}, or chemical characterization of bonding changes after thermal cycling (NMR \cite{SunRef3,SunRef4,SunRef5,SunRef14,SunRef15,SunRef25,SunRef26,SunRef30,SunRef31,SunRef33,SunRef35}, FTIR \cite{SunRef4,SunRef19,SunRef20,SunRef26,SunRef34}, and Raman \cite{SunRef3,SunRef29,SunRef35}). In this work, we exploit the lateral gradient (single-stripe) laser spike annealing (lgLSA) technique \cite{SunRef45} to quantify chemical transformation (bridge cleavage, fully networked framework formation, and template removal) following much shorter timescale sub-millisecond heating. By studying the structural development as a function of the heating duration and peak anneal temperature in this range, we probe the structural transformations and bridge cleavages kinetics, and determine individual reaction activation enthalpies. By understanding the sequence of reactions and their rates, this study identifies new synthesis and/or post-processing routes to control the framework and pore formation while minimizing bridge cleavage.

\section{Results and Discussion}

\subsection{Transformation dynamics}

\begin{figure*}[b!]
\centering
\includegraphics[width= \linewidth]{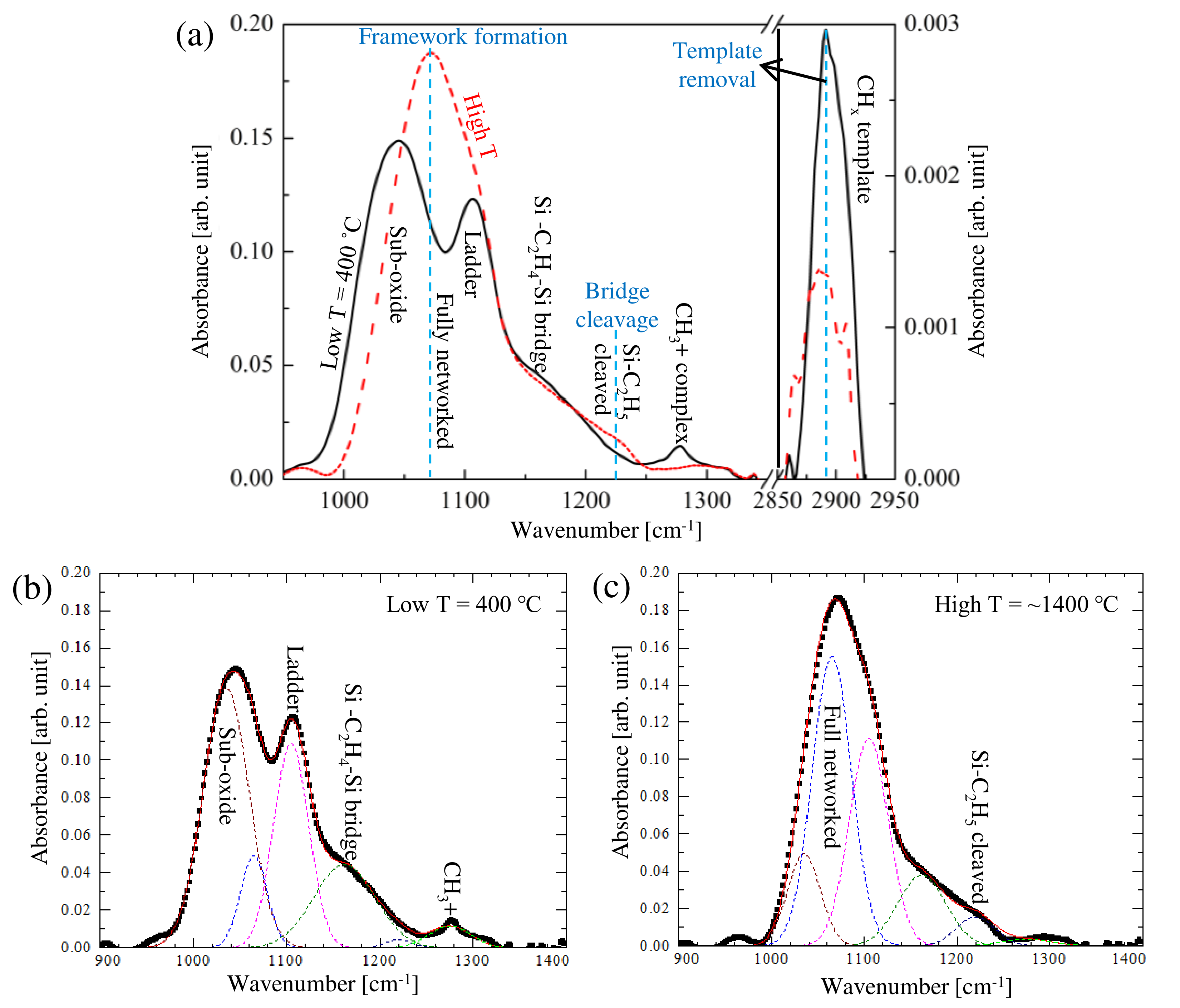}
\caption{(a) Comparison of FTIR spectra after ambient heating to 400\,℃ (``low T") and to $\sim$\,1400\,℃ (``high T") for 3\,ms. (b,c) Gaussian peak fits showing sub-oxide, fully networked, ladder structures, Si-C$_2$H$_4$-Si bridge, Si-C$_2$H$_5$ bridge cleavage, and [CH$_3$+] motifs at 400\,℃ (b) and $\sim$\,1400\,℃ (c).}
\label{SunFig2}
\end{figure*}

FTIR spectra were collected as a function of the peak anneal temperature (up to $\sim$\,1400\,\textdegree C for heating dwells from 0.5 to 3\,ms. \textbf{Figure~\ref{SunFig2}}a compares FTIR data from a sample exposed only to ambient conditions (400\,℃ sample chuck temperature) to one annealed to $\sim$\,1400\,℃ for 3\,ms. Detailed data for all conditions (LSA peak temperatures from 400\,℃ to $\sim$\,1400\,℃; heating dwells of 0.5, 1, 1.5, and 3\,ms) are included in Figure~S1. The fully networked Si-O framework formation, carbon-bridge cleavage, and template removal were characterized. The Si-O FTIR peaks include the fully networked framework at 1072\,cm$^{-1}$, sub-oxide SiO$_x$C$_y$ at 1035\,cm$^{-1}$, and the ladder-like structure at 1105\,cm$^{-1}$. The carbon-bridge cleavage was determined by the change of the Si-CH$_2$-CH$_2$-Si bridge peak at 1162\,cm$^{-1}$ and a newly appearing Si-CH$_2$CH$_3$ ethyl chain peak at 1220\,cm$^{-1}$. Changes in the template species were determined from the CH$_x$ (2860\,--\,2920\,cm$^{-1}$) and CH$_3$+ (1275\,cm$^{-1}$) organic complexes \cite{SunRef40}. These sub-peaks were fit as Gaussians (Figure~\ref{SunFig2}b,c), with integrated areas calculated for quantification.

\bigbreak

The temperature-dependent structural transformations as a function of the laser dwell are presented in Figure~S2~and~S4. The transformation fraction follows the typical sigmoidal behavior with three temperature regimes: (1) an incubation range with minimal transformation; (2) an intermediate range with rapidly increasing transformation fraction, and (3) a saturation regime where the fraction approaches asymptotically 100\% conversion. Depending on the structure’s thermal stability, the temperature ranges of these stages will vary.

\bigbreak

\begin{figure*}[tb!]
\centering
\includegraphics[width= \linewidth]{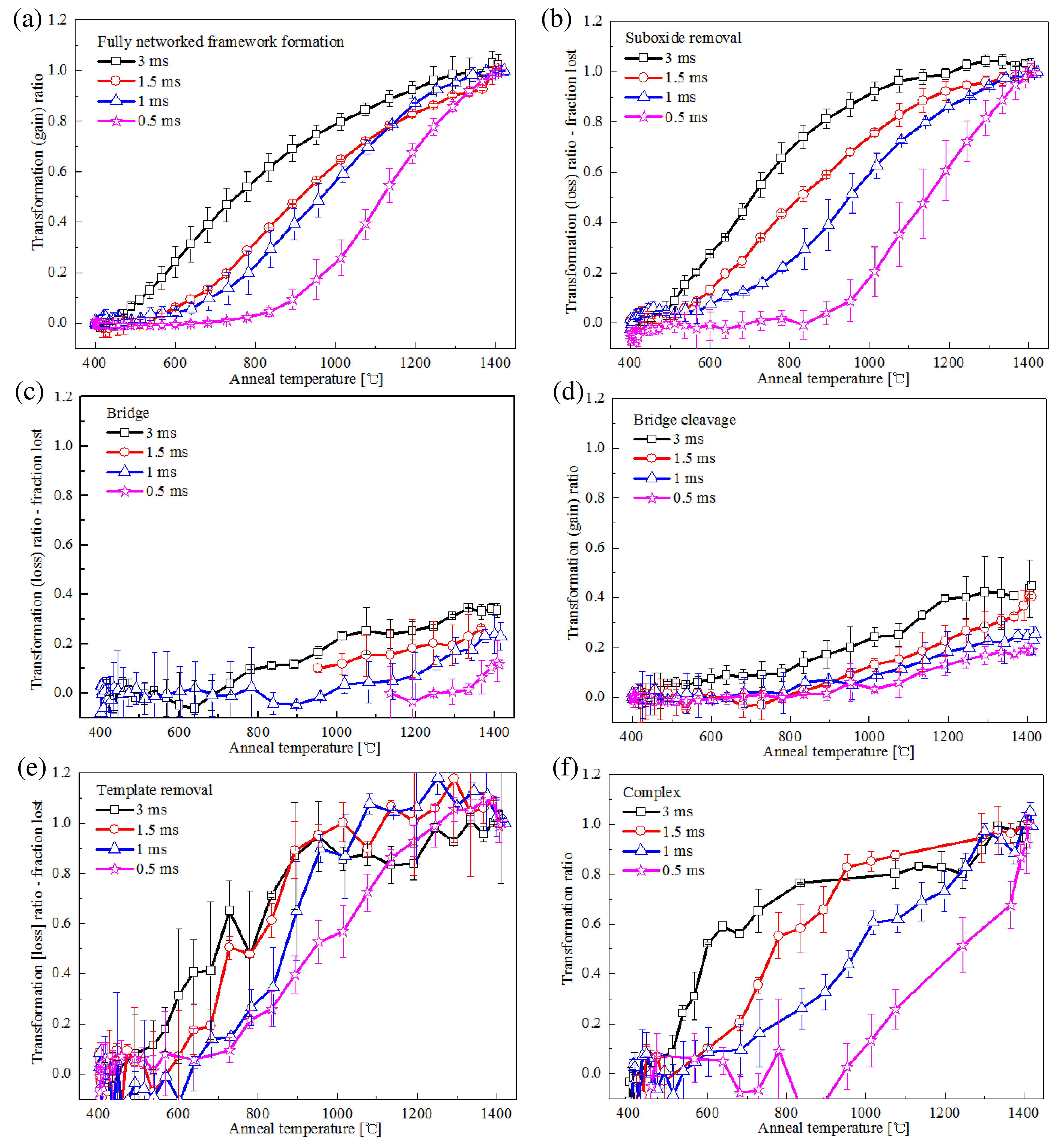}
\caption{Structural transformation dynamics as a function of LSA temperature at different dwells. (a) Fully networked framework formation, (b) sub-oxide removal, (c) bridge (Si-C$_2$H$_4$-Si), (d) bridge cleavage (Si-C$_2$H$_5$), (e) template removal (CH$_x$), and (f) complex [CH$_3$+].}
\label{SunFig3}
\end{figure*}

With increasing temperature, fully networked frameworks (Figure~S2a) form from organic sub-oxides (Figure~S2b). In contrast, the intensity of the ladder-like structure (Figure~S2c) remains constant with temperature and is not analyzed further. To minimize influences from slight differences in the initial concentrations, a transformation ratio was calculated as $\textrm{Ratio}\,=\,(\textrm{Loss}\,\textrm{or}\,\textrm{gain})\,/\,\textrm{Total}.$ The fully networked framework formation (\textbf{Figure~\ref{SunFig3}}a) and sub-oxide removal (Figure~\ref{SunFig3}b) exhibit similar transformation ratios with temperature and time.

\bigbreak

In contrast, a significant fraction of carbon bridges (Si-C$_2$H$_4$-Si peak) remain intact even at the highest temperature and dwell studied (Figure~\ref{SunFig3}c), with the onset of bridge loss shifting to higher temperatures for shorter dwells. For the 0.5\,ms dwell, the bridge peak exhibits minimal changes. 

\bigbreak

The bridge cleavage can also be quantified by the rise of the Si-CH$_2$CH$_3$ peak (Figure~S4b); the fully cleaved intensity was estimated by a long-duration hotplate anneal \cite{SunRef40}. The Si-C$_2$H$_5$ data (Figure~\ref{SunFig3}d) is consistent with the bridge loss data (Figure~\ref{SunFig3}c), confirming the limited loss of bridges for short dwells. At the 0.5\,ms dwell, bridge cleavage is minimal until $\sim$\,1000\,℃, while at 3\,ms dwell the significant loss is observed even by $\sim$\,800\,℃. While the onset of loss occurs earlier for long dwells, the total loss is only 20\% for 0.5\,ms or 40\% for 3\,ms dwells at the highest LSA heating temperature.

\bigbreak

The template removal was quantified using the [CH$_x$] FTIR peak at 2860\,--\,2920\,cm$^{-1}$ (Figure~\ref{SunFig3}e); as other chemical entities also would contribute to this peak, we assume full conversion occurs at the highest temperatures as evidenced by the saturation of this peak above 1000\,℃ (Figure~S4c). The template removal depends primarily on the peak temperature with only weak dependence on the dwell. By 1000\,℃, nearly all the templates are lost for the 1\,--\,3 ms annealed samples, and $>$\,60\% for the 0.5\,ms sample. 

\bigbreak

The [CH$_3$+] organic complexes FTIR peak (Figure~\ref{SunFig3}f) is also related to the template loss but exhibits slightly different behaviors. This peak shows much stronger dwell dependence and mimics more of the network-forming kinetics. This suggests the signal arises from multiple entities including the template, organic SiO$_x$C$_y$ sub-oxide, and carbon bridges.

\subsection{Activation energies}

\begin{figure*}[b!]
\centering
\includegraphics[width= \linewidth]{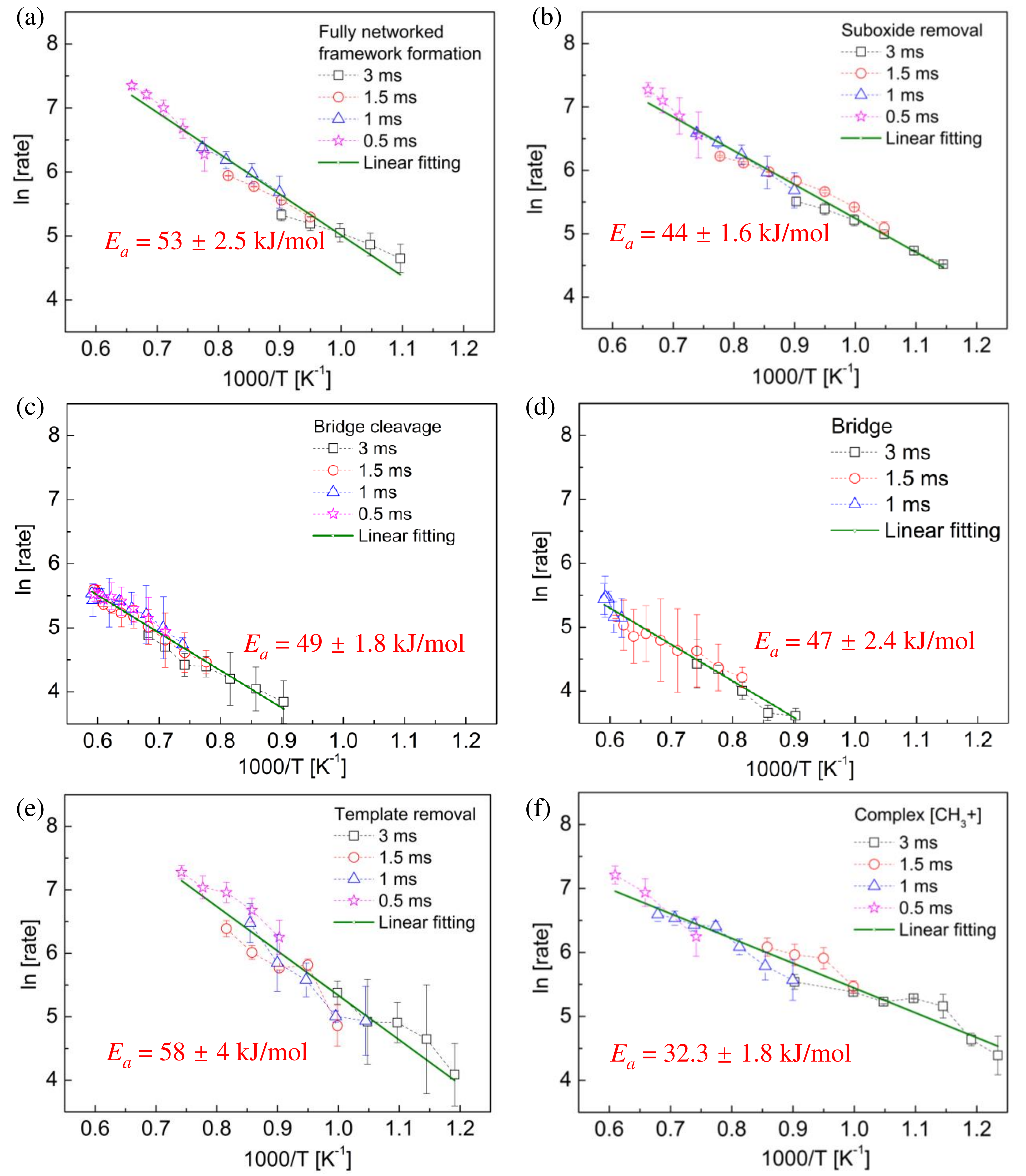}
\caption{Arrhenius plots and activation energies for the structural transformation. (a) Fully networked framework formation, (b) sub-oxide removal, (c) bridge cleavage via Si-CH$_2$CH$_3$ formation, (d) bridge cleavage via loss of Si-CH$_2$CH$_2$-Si, (e) template (porogen) removal (CH$_x$), and (f) organic complex loss [CH$_3$+]. Datasets for bridge cleavage and bridge motifs are taken at a transformation range from 10\% to the highest transformation limit; all other datasets are taken at a fixed transformation range from 25\% to 75\%.}
\label{SunFig4}
\end{figure*}

Taking advantage of an extended temperature range for each dwell, activation energies were extracted from Arrhenius plots of the six structural transformations as shown in \textbf{Figure~\ref{SunFig4}}. Data were fit to the Arrhenius equation, $$\textrm{Rate}\,=\,C\,\mathrm{e}^{-\,E_\mathrm{a}\,/\,R\,T},$$ where $E_{a}$ is the activation enthalpy, $T$ is the LSA heating peak temperature, $R$ is the gas constant (8.314\,J/mol), and $C$ is a constant pre-factor. Activation energies were calculated from the slopes of the log rate versus 1/$T$, with slopes essentially independent of the dwell. \textbf{Table~\ref{tbl:2}} summarizes these critical activation energies.

\bigbreak

We find that template removal and full-network formation require the highest activation energies (50\,--\,62\,kJ/mol), while removal of sub-oxide and complex [CH$_3$+] show the smallest activation energies (30\,--\,46\,kJ/mol). The bridge-cleavage indicators (Si-CH$_2$CH$_3$ and Si-CH$_2$-CH$_2$-Si) exhibit an intermediate range of activation energies (45\,--\,51 kJ/mol). This suggests that removal of large organic molecules and relaxation of the inorganic framework will be more difficult to kinetically activate than the local relaxation of small organic structures. 

\bigbreak

With increasing temperature and correspondingly shorter times, processes with high activation energies (template removal and network formation) will be kinetically accelerated more rapidly than low activation processes (bridge cleavage and sub-oxide removal). This provides a kinetic pathway to optimize overall film development during thermal annealing.   

\begin{table} [htbp!]
    \centering
  \caption{Summary of activation energies ($E_\mathrm{a}$) for different structural transformations.}
  \label{tbl:2}
  \begin{tabular}{p{2.3cm} p{1.8cm} p{1.8cm} p{1.8cm} p{1.8cm} p{1.8cm} p{1.8cm}}
    \hline
    Structural transformation	& Template removal	& Fully networked	& Bridge cleavage	& Bridge	& Sub-oxide removal	& Complex [CH$_3$+] \\
    \hline
    $E_\mathrm{a}$\,[kJ/mol] &	58\,±\,4 &	53\,±\,2.5 &	49\,±\,1.8 &	47\,±\,2.4 &	44\,±\,1.6 &	32\,±\,1.8 \\
   \hline
\end{tabular}
\end{table}

\subsection{Effective local heating time}

\begin{figure*}[b!]
\centering
\includegraphics[width= \linewidth]{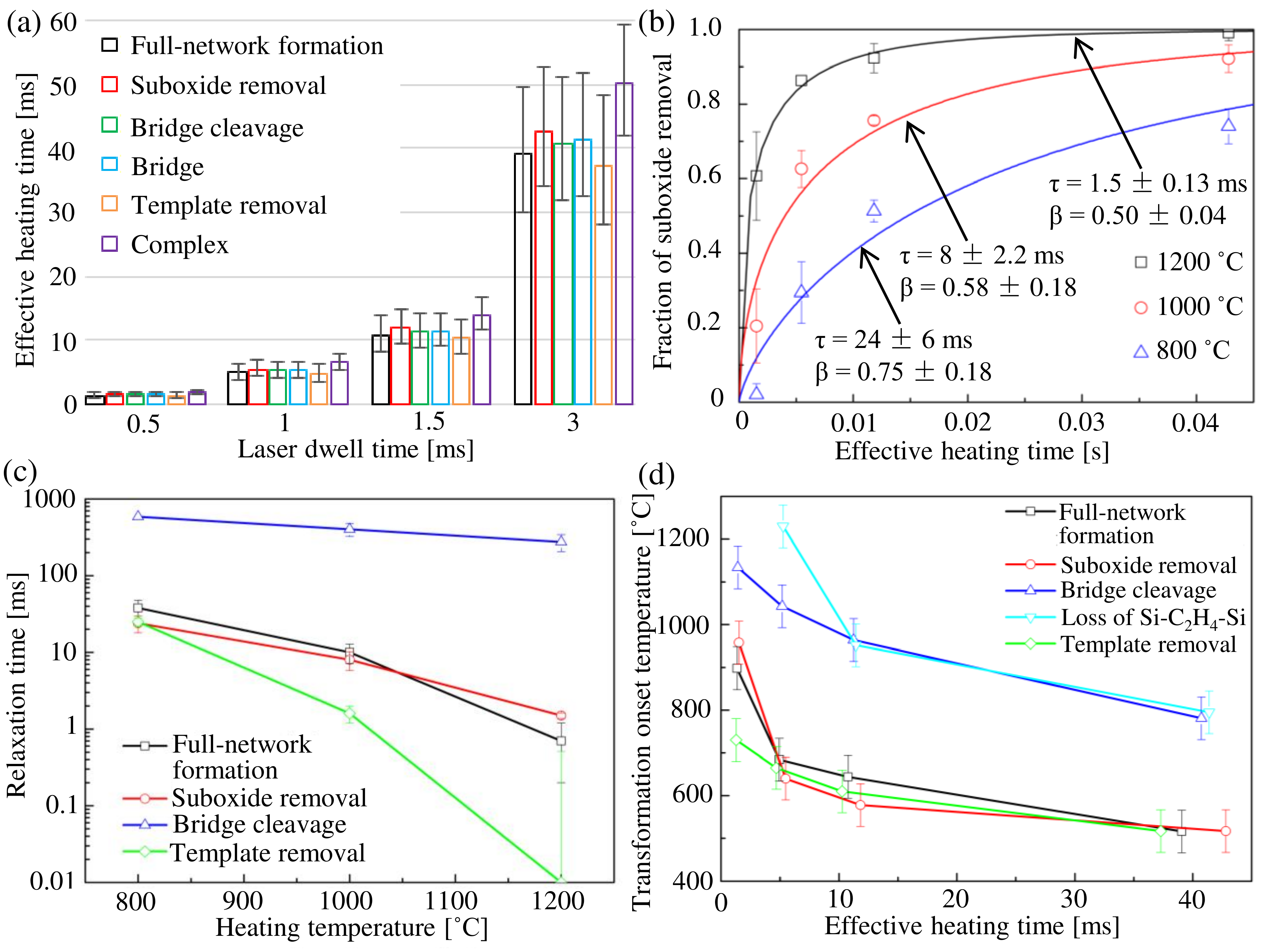}
\caption{Kinetic analysis of structural transformation. (a) Effective local heating time for various laser dwells and structural transformations at $T_\mathrm{max}$$\,\simeq\,$1400\,\textdegree C. (b) Example of fitting the stretched exponential for sub-oxide removal. (c) Relaxation time for transformations as a function of temperature. (d) Onset temperature for 10\% transformations as a function of heating duration.}
\label{SunFig5}
\end{figure*}

In contrast to isothermal anneals, LSA temperatures vary rapidly with time and are characterized by a dwell time ($\tau_\mathrm{dwell}$). Using the determined activation energies, the effective time at the peak temperature ($T_\mathrm{max}$)  is given by $$ t_\mathrm{eff}\,=\,\int\exp\left\lbrace-\,\frac{E_\mathrm{a}}{k}\,\left[\frac{1}{T(t)}\,-\,\frac{1}{T_\mathrm{max}}\right]\right\rbrace\,\mathrm{d}t,$$ where $E_\mathrm{a}$ is the activation energy, $\mathit{k}$ is the Boltzmann constant, and $T(t)$ is the time-dependent temperature. The time dependence $T(t)$ was determined from heat-flow simulations using CLASP \cite{SunRef62}. As shown in \textbf{Figure~\ref{SunFig5}}a, different laser dwells result in a distinctive variation in effective heating duration. For a 3\,ms dwell, the effective time is $\sim$\,40\,ms, while 0.5\,ms dwell samples experienced $\sim$\,1.5\,ms effective heating, with small variations due to the different activation energies. The increase in $t_\mathrm{eff}\,/\,\tau_\mathrm{dwell}$ with increasing dwell arises from the transition from a thermally thick substrate (fast quench) to a thermally thin substrate (slow quench). 

\subsection{Stretched exponential relaxation}

While the bridge-cleavage activation energy is not significantly higher than that for the template removal (Table~\ref{tbl:2}), the structural dynamic data (Figure~\ref{SunFig3}c,d) demonstrates that bridge cleavage is substantially minimized under sub-millisecond heating. Understanding this behavior requires determining the fundamental processes that control structural transformations in PMOs and their detailed kinetic rates.

\bigbreak

Structural transformations require a continuous supply of reactive species that diffuse from the bulk surface, or that are generated within the film. For disordered transformations with a range of activation energies, the stretched exponential is commonly used to model transformation kinetics \cite{SunRef47,SunRef48,SunRef49,SunRef50,SunRef52}. The PMO structure contains both ordered and periodic features (pores separated by Si-O based wall as shown in Figure~\ref{SunFig1}),  with carbon-bridge motifs randomly inserted in the networked wall \cite{SunRef6,SunRef8,SunRef9,SunRef10}. This complex system is expected to follow a diffusion-controlled stretched exponential expression featuring random-distributed, time-dependent diffusion. 

\bigbreak

Transformation dynamics were fit by a stretched exponential (Figure~\ref{SunFig5}b) $$X\,=\,1\,-\,\mathrm{exp}\,\left[\,-\,\Bigl(\frac{t}{\tau}\Bigl)^{\beta}\,\right],$$ where $\mathit{X}$ is the transformation ratio, $\mathit{t}$ is the time, $\tau$ is a characteristic time constant (required relaxation time), and $\beta$ is an index that reflects coupling of relaxation processes. Details of fitting for different motifs are included in Figure~S5, together with the summarized parameters in Table~S1~and~S2. 

\bigbreak

The relaxation time constants for structural transformations are presented in Figure~\ref{SunFig5}c as a function of the peak temperature. The time to induce bridge cleavage (200\,--\,600\,ms) is significantly higher than that for framework formation, sub-oxide removal, and template removal (1\,--\,50 ms). For example, at 1200\,℃, the relaxation time for bridge cleavage is $\sim$\,300\,ms, while for framework formation and template removal, it is only $\sim$\,1 ms. This large difference in relaxation times enables the manipulation of different structural modifications using the time and temperature trade-off. This behavior reflects on the essentially higher onset temperatures for bridge cleavage than other transformations (Figure~\ref{SunFig5}d).

\bigbreak

\textbf{Table~\ref{tbl:3}} compares the ratio of effective heating at 1200\,℃ for 0.5\,ms and 3\,ms dwells to the relaxation time constant. Even at 1200\,℃ for 3\,ms, there is insufficient time to significantly impact the bridge cleavage ($t_\mathrm{eff}$\,/\,$\tau_\mathrm{relax}$\,=\,0.15\,±\,0.3), while for 0.5\,ms there is already sufficient time for complete network formation and template/sub-oxide removal.

\begin{table} [htbp]
  \caption{Comparison of the ratio of the effective heating time to the structural relaxation time at 1200\,℃ for 0.5\,ms and 3\,ms dwells.}
  \centering
  \label{tbl:3}
  \begin{tabular}{p{1.9cm} p{3cm} p{3cm} p{3cm} p{3cm}}
    \hline
    Laser dwell	& Bridge cleavage & Sub-oxide removal & Fully networked & Template removal \\
    \hline
    0.5\,ms &	0.005\,±\,0.3	 & 1\,±\,0.22 & 2\,±\,0.7 &	$>$\,30 \\
    3\,ms & 0.15\,±\,0.3 & 29\,±\,0.21 & 56\,±\,0.7	& $>$\,800	 \\
   \hline
\end{tabular}
\end{table}

\subsection{Transformation diagram and heat treatment design}

\begin{figure*}[b!]
\centering
\includegraphics[width= \linewidth]{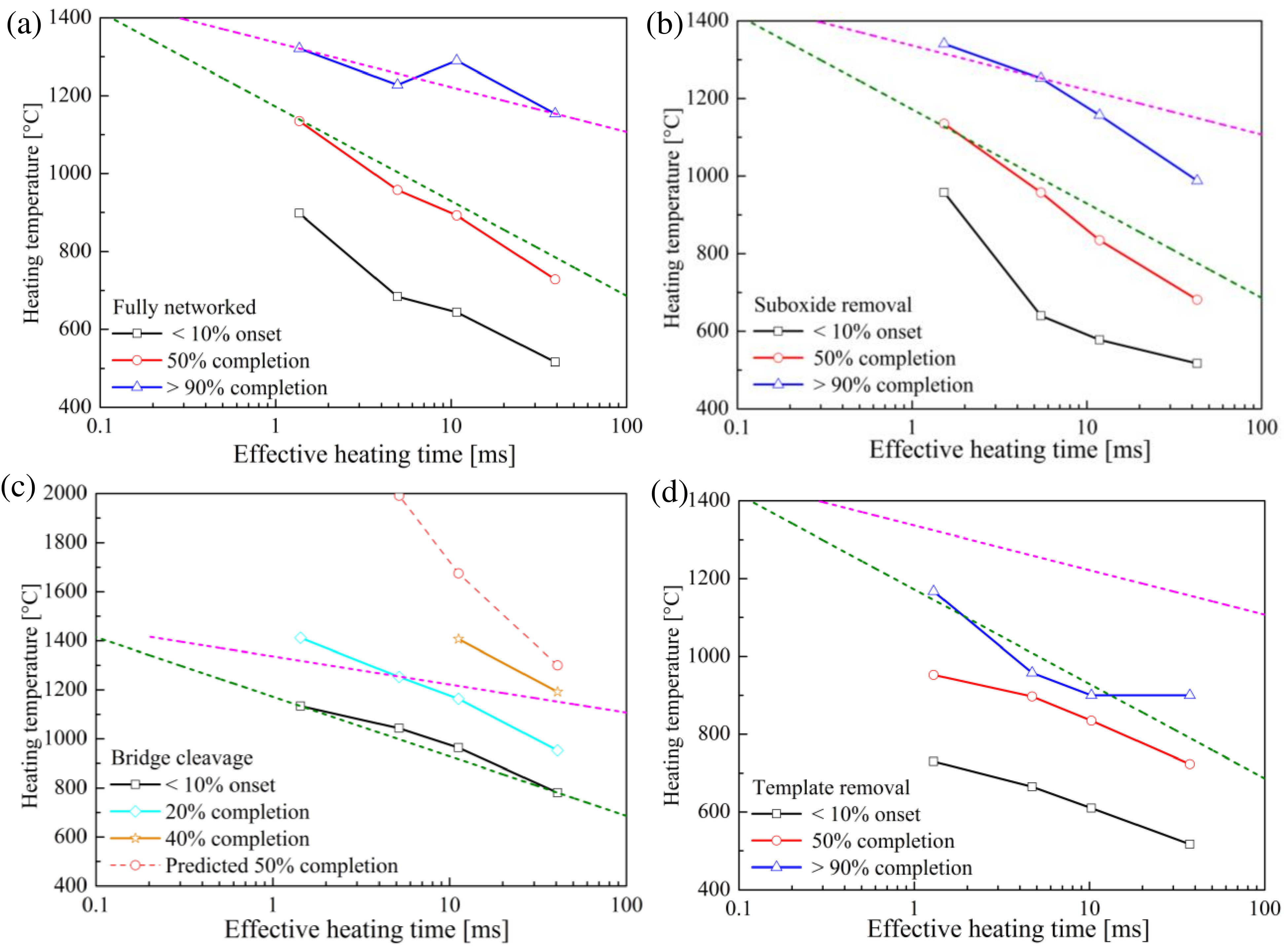}
\caption{Structural transformation diagram showing critical points during different transformation stages. (a) framework formation, (b) sub-oxide removal, (c) bridge cleavage (Si-C$_2$H$_5$), (d) template removal (CH$_x$). The green dash line represents the maximum temperature at each dwell that can be tolerated with minimized loss of the bridge motif (set by 6c). The purple dash line represents the minimum temperature at each dwell that fully transforms the required framework and template (set by 6a).}
\label{SunFig6}
\end{figure*}

In order to facilitate PMO applications, the structural transformation diagram with critical heating temperature versus time is shown in \textbf{Figure~\ref{SunFig6}}. In these diagrams, the green curve represents the maximum temperature that can be tolerated with minimal loss ($<$\,10\%) of the bridge motif. For each motif, except the bridge, the onset ($<$\,10\%), completion ($>$\,90\%), and 50\% transformation curves are shown. For the bridge cleavage, the 20\% and 40\% transformation curves are shown with an estimate for the temperature that would be required to achieve 50\% transformation. To avoid significant loss of bridge structures, the suboxide-to-network conversion is limited to $\sim$\,50\% while the template conversion is nearly completed. At the other extreme, the purple curves show the minimum temperature required as a function of heating time to fully convert the network. Loss of bridge motif is minimized by the shortest heating times with less than 20\% loss for $t_\mathrm{heat}$\,=\,1\,ms.   

\section{Conclusion}

In summary, the annealing kinetics of bridged PMOs were investigated as a function of LSA peak temperatures (400\,--\,1400\,\textdegree C) and dwells (0.5\,--\,3\,ms). We established the transformation dynamics of bridge cleavage, framework formation, and template removal, with activation energies of $58\pm 4~\mbox{kJ/mol}$ for template removal, $53\pm 2.5$ for formation of the full network, $48\pm 2$ for bridge loss, and $44\pm 1.6$ for removal of sub-oxide structures. 

\bigbreak

While the activation energies for the various processes are similar, the kinetics vary significantly quantified by a relaxation time.  At 1200\,\textdegree C, the relaxation times are 300\,ms for bridge cleavage but only $\approx$\,1\,ms for framework formation and template removal.  This large difference in relaxation times allows network formation to be completed while the bridge cleavage is minimized under heating at sufficiently high temperature and short times. 

\bigbreak

In applying this finding to heat treatment design, a time-temperature-transformation diagram is developed. Effects of porosity and reaction routes (see Supplementary Information) could be further studied using this same methodology. The control of framework and pore formation without bridge cleavage could enable new synthesis and post processing, and facilitate the transition of PMO-class materials to broad applications. 

\section{Experimental Section}

\threesubsection{Sol-gel PMO-film synthesis} 

1,2-Bis(triethoxysilyl)ethane (BTEE, 96\%), Brij-76, and 2-methoxy-1-propanol (PMOH, $>$\,99.5\%) were purchased from Sigma-Aldrich and used as received without further purification. 
\bigbreak
Ethyl-bridged periodic mesoporous organosilicas (PMOs) were synthesized in PMOH solution at room temperature through a template-directed sol-gel process. The precursor BTEE and template (porogen) Brij-76 were diluted separately in PMOH to 25\,wt\% concentration. The template was added to the BTEE solution at 9\,--\,21.5 wt\% ratios, and then 0.61\,mL nitric acid (1\,M) was added per gram of BTEE. The solution was aged for 15 minutes. 
\bigbreak
PMO thin films were prepared by spin-coating on nominally undoped bare Si wafers at 2000\,RPM for 90 seconds. Films were immediately baked at 85\,\textdegree C for 2 minutes on a hotplate, followed by pre-cure baking at 400\,\textdegree C for 1 hour in a vacuum box oven (Yield Engineering Systems, 450PB). This pre-cure allowed the initialization of sol-gel reactions and provided an initial porous environment with the necessary frameworks for the high-temperature LSA investigations. After the pre-cure bake, films were $\sim$\,300\,nm thick.   
\bigbreak
The sol-gel hydrolysis and condensation reactions of precursors and templates (porogen) are given in Figure~S9, and their intermediate structures, as determined by FTIR, are shown in Figure~S10. Figure~S11 shows the ideal chemical structure of a resulting film with intrinsic carbon bridges embedded in the Si-O network. 

\threesubsection{Characterization} 

Chemical structure changes were determined using transmission FTIR (Fourier-transform infrared spectroscopy, Bruker Hyperion) and XPS (X-ray photoelectron spectroscopy). In order to achieve spatially resolved characterization, FTIR measurements were obtained with a 20\,$\mu$m aperture scanned across the $\sim$\,700\,$\mu$m laser-annealed lines with spectra every 25\,$\mu$m. The FTIR spectra were fit to nine Gaussian peaks (in four sets): precursor peaks of Si-OH at 900\,cm$^{-1}$ and Si-OCH$_2$CH$_3$ at 930\,cm$^{-1}$; Si-O framework peaks of sub-oxide at 1023\,--\,1035\,cm$^{-1}$, fully networked Si-O framework at 1065\,--\,1072\,cm$^{-1}$, and ladder structure at 1105\,cm$^{-1}$; carbon-bridge related peaks of Si-CH$_2$-CH$_2$-Si at 1150\,--\,1162\,cm$^{-1}$ and Si-CH$_2$CH$_3$ (bridge cleavage) at 1220\,cm$^{-1}$; and template (porogen) peaks of complex CH$_x$ at 2860\,--\,2920\,cm$^{-1}$ and CH$_3$+ at 1272\,--\,1275\,cm$^{-1}$. Details of the fitting procedure have been published previously \cite{SunRef40}, with detailed analyses of the sol-gel reactions after deposition and pre-cure given in the Supplementary Information. 

\threesubsection{Laser spike annealing (LSA)} 

Following pre-cure, films were annealed to high temperatures using the lateral gradient (single-stripe) LSA (lgLSA) technique \cite{SunRef45}. A 120\,W CO$_2$ laser ($\lambda$\,=\,10.6\,$\mu$m) was focused to a line-shape beam ($\sim$\,95\,$\mu$m\,$\times$\,$\sim$\,700\,$\mu$m) and scanned across films/substrates mounted on a 400\,℃ pre-heated vacuum chuck. The lateral intensity of the laser was intentionally near Gaussian to establish a lateral temperature profile across laser scans. Dwell time, defined as the laser FWHM in the scan direction divided by the scan velocity, was varied between 0.5 and 3\,ms. Peak temperature at the center of each stripe was adjusted by the incident laser power using an optical attenuator, with peak temperatures for 400\,\textdegree C (hot stage) to $\sim$\,1400\,\textdegree C (Si melt). Temperatures were calibrated using Si and Au melt coupled with CLASP heat flow simulations \cite{SunRef62}.  

\medskip
\textbf{Supporting Information} \par
Supporting Information is in another file, including FTIR and XPS analyses on sol-gel samples, details on structural dynamics and stretched exponential fitting; discussions on the effects of porosity and on possible reaction routes in PMO annealing. 

\medskip
\textbf{Acknowledgements} \par 

The work was supported by the Semiconductor Research Corporation (SRC), GlobalFoundries, and Intel. This work was performed in part at the Cornell NanoScale Facility, an NNCI member supported by NSF Grant NNCI-2025233, and made use of the Cornell Center for Materials Research Shared Facilities which are supported through the NSF MRSEC program (DMR-1719875). Z.S. acknowledges Dr. Y. Sun and Dr. D. Zhang for experimental assistance. 

\medskip
\textbf{Author Contributions} \par 

Z.S. performed the experiments, characterizations, and analyses. A.C. assisted with laser spike annealing and thermal analysis. M.O.T. supervised the work and analyzed the data. Z.S. and M.O.T. wrote the manuscript. 

\medskip
\textbf{Conflicts of Interest} \par 

The authors declare no competing financial interests.

\medskip
\textbf{Data Availability Statement} \par 

Data are available by contacting the corresponding author.

\end{sloppypar}
\end{document}